\documentclass[aps, prb, twocolumn,amsmath,amssymb,floatfix, reprint, longbibliography]{revtex4-2}
\usepackage{graphicx}
\usepackage{svg}
\usepackage{times}
\usepackage{dcolumn}
\usepackage{hyperref}
\usepackage{chemformula}
\hypersetup{colorlinks, allcolors=blue}
\usepackage{lipsum}
\usepackage{comment}
\newcommand{\kk}{{\mathbf{k}}}
\newcommand{\kp}{{\mathbf{k}^\prime}}
\newcommand{\qq}{{\mathbf{q}}}

\begin{document}
\title{Constrained weak-coupling superconductivity in  multiband superconductors}
\author{Niels Henrik Aase}
\email[These authors contributed equally to this work]{}
\affiliation{\mbox{Center for Quantum Spintronics, Department of Physics, Norwegian University of Science and Technology, NO-7491 Trondheim, Norway}}
\author{Christian Svingen Johnsen}
\email[These authors contributed equally to this work]{}
\affiliation{\mbox{Center for Quantum Spintronics, Department of Physics, Norwegian University of Science and Technology, NO-7491 Trondheim, Norway}}
\author{Asle Sudb{\o}}
\email[Corresponding author: ]{asle.sudbo@ntnu.no}
\affiliation{\mbox{Center for Quantum Spintronics, Department of Physics, Norwegian University of Science and Technology, NO-7491 Trondheim, Norway}}

\begin{abstract}
We consider superconductivity in a system with $N$ Fermi surfaces, including intraband and interband effective electron-electron interactions. The effective interaction is described by an $N \times N$ matrix whose elements are assumed to be constant in thin momentum shells around each Fermi surface, giving rise to $s$-wave superconductivity. Starting with attractive intraband interactions in all $N$ bands, we show that too strong interband interactions are detrimental to sustaining $N$ nonzero components of the superconducting order parameter.
We find similar results in systems with repulsive intraband interactions. 
The dimensionality reduction of the order-parameter space is given by the number of nonpositive eigenvalues of the interaction matrix. 
Using general models and models for superconducting transition metal dichalcogenides and iron pnictides frequently employed in the literature, we show that constraints must be imposed on the order parameter to ensure a lower bound on the free energy and that subsequent higher-order expansions around the global minimum are thermodynamically stable.
We also demonstrate that similar considerations are necessary for unconventional pairing symmetries.
\end{abstract}

\maketitle
\section{Introduction}
Superconductivity in systems with multisheeted Fermi surfaces was first studied \cite{Suhl1959} shortly after the introduction of the basic Bardeen-Cooper-Schrieffer (BCS) theory \cite{Bardeen1957} of weak-coupling superconductivity. More recently, superconductivity has been discovered in various compounds with multiple bands crossing the Fermi surface, such as \ch{MgB_2} \cite{Nagamatsu2001}, iron pnictides \cite{Kamihara2006, Kamihara2008}, and monolayer transition metal dichalgocentides (TMDs) such as \ch{NbSe_2} \cite{Xi2016} or \ch{MoS_2} \cite{Lu2015}. From a theoretical point of view, these systems are of interest due to a number of phenomena that are predicted \cite{Milosevic2015}, with no counterparts in one-band superconductors. One such feature is multiphase physics involving the individual phases of the components of the superconducting order parameter. This may lead to spontaneously broken time-reversal symmetry \cite{Ng2009, Stanev2010, Tanaka2010, Carlstrom2011, Lin2012, Takahashi2014, Bojesen2014} and various types of phase fluctuations such as Leggett modes \cite{Leggett1966}. Multiband systems have also gained significant attention due to their potential for unconventional pairing \cite{Kuroki2008, Hirschfeld2011, Scalapino2012, Chubukov2012}, higher critical temperatures \cite{Perali2013, Uebelacker2012, Trevisan2018, Shang2019, Salasnich2019}, improved performance in magnetic fields \cite{Kuzmanovi2022, Ghanbari2022, Salamone2023}, and potential for topological superconductivity \cite{Yuan2014, He2018, Shaffer2020}. 

The BCS theory is a cornerstone of superconductivity research. In its original form, it was applied to systems with only a single electron band crossing the Fermi surface. Given its relative simplicity compared to other more elaborate theories applicable in the strong-coupling regime, the extension of BCS theory to multiband systems is a natural choice for treating the abovementioned systems, but still with the limitations that come from the weak-coupling approach. In the simplest one-band case, which yields $s$-wave superconductivity, a necessary and sufficient condition for finding a thermodynamically stable superconducting state is that electrons interact via an {\it attractive} effective interaction operative in a thin momentum shell around the Fermi surface. The effective interaction is taken to be a constant within this thin momentum shell. 

In multiband superconductors with electron pairing possible on each band crossing the Fermi surface and with corresponding $s$-wave multicomponent superconducting order parameters, the situation is more complicated. The BCS type of constant interaction operative within thin momentum shells of the Fermi surface is now a matrix whose elements describe intraband as well as interband scattering of electrons. Nontrivial solutions to the equations determining the superconducting order parameter, describing thermodynamically stable superconducting states, may then be found even if some of the matrix elements represent repulsive interactions. Since $s$-wave Cooper pairing of electrons somehow requires attractive interaction between electrons, such attraction from repulsion represents a fascinating possible aspect of multiband superconductivity. 

To study these $N$-band systems, a set of $N$ coupled equations for the complex components of the superconducting order parameter (gap equations) are frequently employed in the literature. They were originally derived as self-consistency relations \cite{Suhl1959} and stationary-point conditions on the free energy of the superconducting state \cite{Kondo1963}. Some examples include studies of the time-reversal symmetry breaking $s + is$ states in iron pnictides with some repulsive interaction-matrix elements \cite{Tanaka2010,Stanev2010,Maiti2013,Stanev2012, Wilson2013}, and more recent investigations of Ising superconductivity in the TMD \ch{NbSe_2} \cite{Horhold2023,Das2023}. Multiband gap equations have recently also been found to be relevant in single-band systems subject to magnetic fields \cite{Shaffer2021, Shaffer2022}.

One complicating factor in the multiband case is that the criterion for finding the superconducting order parameter that represents a {\it thermodynamically stable superconducting state} is less obvious than in the one-band case. The effective electron interactions are contained in an interaction matrix, and the simple criterion for the solutions to represent a thermodynamically stable superconducting state for the one-band case is no longer applicable. Furthermore, since the order parameter has $N$ complex components in the $N$-band case, the possibility of multiple solutions to the gap equations needs to be considered seriously. This becomes increasingly complicated with increasing $N$. Remarkably, in the one-band case, a nontrivial solution to the equation for the superconducting order parameter always describes the thermodynamically stable state. On the other hand, in the multiband case, one is faced with the problem of deciding which solution, if any, describes an actual stable superconducting state. That the situation might call for more care than in the one-band case has been briefly noted in previous papers on multiband superconductivity \cite{Stanev2012, Iskin2016}. 

The standard procedure for finding a superconducting ground state is to check that the normal-state free energy is higher than the superconducting free energy, and in the case of multiple solutions to the gap equations, the one with the lowest energy is chosen.
In this paper, we point out that this procedure is not always sufficient, since the free energy of the superconducting state may be \emph{unbounded from below} for a wide class of multiband systems. In such cases, we outline a method that allows for capturing the relevant physics, in spite of the seemingly nonphysical theory.

We consider multiband $s$-wave superconductors within the weak-coupling BCS approach, focusing on the necessary conditions on the effective electron interaction matrix for nontrivial solutions to the gap equations to represent a thermodynamically stable superconducting state. We demonstrate that in several multiband cases, nontrivial solutions to the gap equations may be found that represent unphysical states, some of which have lower free energy than the normal state. In the order-parameter space of multicomponent $s$-wave superconductors, these states represent either saddle points or maxima in the free energy. As such, the solutions satisfy stationary-point conditions on the free energy, but do not correspond to stable or metastable superconducting states. We elucidate the origin of this behavior and show that if the free energy is unbounded from below, care has to be taken to appropriately reduce the dimensionality of the superconducting order-parameter space in order to describe physically meaningful superconducting states. This dimensionality reduction is crucial as it ensures a lower bound on the free energy, allowing higher-order expansions around solutions to the gap equations.

\section{General mean-field formalism}
We consider a multiband model for superconductivity where it is assumed that electrons can form Cooper pairs in each band, such that the effective Hamiltonian is given by
\begin{equation}
    H  = \sum_{{\bf k},\alpha,\sigma}
    \left( \varepsilon_{\kk \alpha} - \mu \right)
    c^{\dagger}_{{\bf k}\alpha\sigma} c_{{\bf k}\alpha\sigma}
    + \sum_{{\bf k},{\bf k}^{\prime},\alpha,\beta}
    V^{\alpha \beta}_{{\bf k}^{\prime}{\bf k}}
    P^{\dagger}_{{\bf k}^{\prime}\alpha} ~ P_{{\bf k}\beta}.
\label{eq:NbandBCS}
\end{equation}
Here, $c^{\dagger}_{\lambda},c_{\lambda}$ are creation and destruction operators of electron states labeled by $\lambda =({\bf k},\alpha,\sigma)$, where ${\bf k}$ is momentum, $\alpha \in \{1, \dots, N\}$ is a band index, $\mu$ is the chemical potential, and $\sigma$ is a spin quantum number. 
The interaction matrix $V^{\alpha \beta}_{{\bf k}^{\prime}{\bf k}}$ scatters electron pairs in state
$\left[ ({\bf k},\beta,\uparrow),(-{\bf k},\beta,\downarrow) \right]$ to state $\left[ ({\bf k}^{\prime},\alpha,\uparrow),(-{\bf k}^{\prime},\alpha,\downarrow) \right]$ with momentum transfer ${\bf q} = {\bf k}^{\prime}-{\bf k}$.
Although we do not specify the microscopic origin of $V^{\alpha \beta}_{{\bf k}^{\prime}{\bf k}}$, it is assumed that electron spins are individually conserved in the process. This is true for instance if $V^{\alpha \beta}_{{\bf k}^{\prime}{\bf k}}$ originates with electron-phonon couplings or electron-magnon couplings when the magnons arise out of collinear spin ground states \cite{Maeland2023}. 
Furthermore, we have defined the pair-creation operator 
$P^\dagger_{{\bf k} \alpha} \equiv 
c^{\dagger}_{-{\bf k}\alpha\downarrow} c^\dagger_{{\bf k}\alpha\uparrow}$
with the corresponding pair-destruction operator given by 
$P_{{\bf k}\alpha} \equiv 
c_{{\bf k}\alpha\uparrow} c_{-{\bf k}\alpha\downarrow}$. In writing down Eq.\ \eqref{eq:NbandBCS}, we have assumed that only electrons of opposite momenta pair and neglected considering Cooper pairs with finite momentum. All cases we will consider in this paper, illustrated in Fig. \ref{fig:fermi_surfaces}, fall within this category, provided a suitable rewriting of the Hamiltonian is performed. We perform a mean-field approximation in the standard way and define the quantities 
$b_{{\bf k}\alpha} \equiv \langle P_{{\bf k} \alpha}\rangle$ and $b^\dagger_{{\bf k}\alpha} \equiv \langle P^\dagger_{{\bf k} \alpha}\rangle$ along with the quantities
\begin{align}
\begin{split}
\Delta_{{\bf k}^{\prime}\alpha} &\equiv  
-\sum_{{\bf k},\beta} V^{\alpha \beta}_{{\bf k}^{\prime}{\bf k}} ~  b_{{\bf k}\beta}, \\
\Delta^\dagger_{{\bf k}\beta} &\equiv 
-\sum_{{\bf k}^{\prime},\alpha} V^{\alpha \beta}_{{\bf k}^{\prime}{\bf k}} ~  b^{\dagger}_{{\bf k}^{\prime}\alpha}. 
\label{eq:Gaps}
\end{split}
\end{align}
With these definitions, we obtain a mean-field Hamiltonian given by 
\begin{align}
\begin{split}
H &= \sum_{{\bf k},\alpha,\sigma}
\left( \varepsilon_{\kk \alpha} - \mu \right)
c^{\dagger}_{{\bf k}\alpha\sigma} c_{{\bf k}\alpha\sigma} \\
&- \sum_{{\bf k}, \alpha}
\left(\Delta^\dagger_{{\bf k}\alpha}  P_{{\bf k} \alpha} 
 + \Delta_{{\bf k}\alpha}  P^\dagger_{{\bf k} \alpha} - \Delta^\dagger_{{\bf k} \alpha} ~ b_{{\bf k}\alpha} \right).
 \label{eq:N-band_MFT_BCS_before_rewrite}
\end{split}
\end{align}
In deriving Eq.\ \eqref{eq:N-band_MFT_BCS_before_rewrite}, we have only included Cooper pairs where both electrons belong to the same band, i.e.\ intraband pairing. Crossband pairing \cite{Vargas-Paredes2020} has been neglected, since we will focus on the common and most robust case of zero-momentum Cooper pairs. However, within the intraband-pairing approximation, we do consider both intraband and interband  \emph{interactions}. Intraband interactions describe the scattering of an electron pair from a two-particle state in one band to a two-particle state in the same band $V^{\alpha \beta}_{{\bf k}^{\prime}{\bf k}}$ with $\alpha=\beta$. Interband interactions describe scattering of an electron pair from one band to a different band, $\alpha \neq \beta$.

In general, using a complete and orthonormalized set of basis functions $\left \{ g_\eta({\bf k}) \right \}$ appropriate for the point-group symmetry of the system, we can write
\begin{eqnarray}
V^{\alpha \beta}_{{\bf k}^{\prime}{\bf k}}
= \sum_{\eta,\eta^{\prime}} \Lambda^{\alpha \beta}_{\eta\eta^{{\prime}}} ~ g_\eta({\bf k}) ~ g_{\eta^{\prime}}({\bf k}^{\prime}), 
\label{eq:interaction}
\end{eqnarray}
with $\Lambda^{\alpha \beta}_{\eta \eta^{\prime}}
= \sum_{{\bf k},{\bf k}^{\prime}} 
V^{\alpha \beta}_{{\bf k}^{\prime}{\bf k}}
~ g_\eta({\bf k}) ~ g_{\eta^{\prime}}({\bf k}^{\prime})$. A subset of the functions $\left \{ g_\eta({\bf k}) \right \}$ will transform as the identity under the operations of the point group of the system, and these are the ones we will focus on. Considering only the simplest ones, $g_0({\bf k}) = \mathrm{const.}$, then $V^{\alpha \beta}_{{\bf k}^{\prime}{\bf k}} = \Lambda^{\alpha \beta}_{00}$. This interaction will be assumed operative within a thin momentum shell of each Fermi surface in bands 
$\varepsilon_{\kk \alpha}$ and $\varepsilon_{\kk \beta}$. Defining $V_{\alpha \beta} = -  \Lambda^{\alpha \beta}_{00}$ we find from Eqs. \eqref{eq:Gaps} that 
$\Delta_{{\bf k}\alpha}$ will be ${\bf k}$ independent, $\Delta_{{\bf k}\alpha} = \Delta_{\alpha}$. In the following, we will assume that $V_{\alpha \beta} = V_{\beta \alpha}$ (and hence, the inverse matrix is also symmetric). We have that 
$\Delta_\alpha = \sum_{{\bf k}, \beta} V_{\alpha \beta} ~ b_{{\bf k}\beta} 
= \sum_{\beta} V_{\alpha \beta} ~ b_\beta$, where we have defined 
$b_\beta = \sum_{\bf k} b_{{\bf k}\beta}$. Provided that $V_{\alpha \beta}$ is invertible, we then have $b_{\alpha} = \sum_{\beta} V^{-1}_{\alpha \beta} \Delta_\beta$. Using all of this in Eq.\ \eqref{eq:N-band_MFT_BCS}, the mean-field Hamiltonian takes the form
\begin{align}
\begin{split}
H &= 
\sum_{\alpha, \beta}\Delta_\alpha^\dagger V_{\alpha\beta}^{-1}\Delta_\beta
+ \sum_{{\bf k},\alpha,\sigma}
\left( \varepsilon_{\kk \alpha} - \mu \right)
c^{\dagger}_{{\bf k}\alpha\sigma} c_{{\bf k}\alpha\sigma}  \\
&- \sum_{{\bf k}, \alpha}
\left(\Delta^\dagger_{\alpha}  P_{{\bf k} \alpha} 
 + \Delta_{\alpha}  P^\dagger_{{\bf k} \alpha} \right).
 \label{eq:N-band_MFT_BCS}
\end{split}
\end{align}
Here, we have neglected terms 
${\cal{O}} \left( \delta b_{{\bf k} \alpha} \delta b^\dagger_{{\bf k} \alpha} \right)$,
where $\delta b_{{\bf k} \alpha} \equiv P_{{\bf k} \alpha} - b_{{\bf k} \alpha}$.
Performing a diagonalization procedure for the fermionic part for each band individually (the above Hamiltonian is diagonal in band indices) and computing the fermion trace, we obtain the free energy of the system at temperature $T$ as follows:
\begin{align}
    \begin{split}
    F(T)&= \sum_{\alpha, \beta}\Delta_\alpha^\dagger V_{\alpha\beta}^{-1}\Delta_\beta + \sum_{\kk,\alpha} \left( \varepsilon_{\kk \alpha} - \mu \right) \\
    &-\frac{1}{\beta}\sum_{\kk,\alpha}\left [\ln(1 + \mathrm{e}^{-\beta E_{\kk \alpha}}) + \ln(1 + \mathrm{e}^{\beta E_{\kk \alpha}})\right],
    \label{F_T}
\end{split}
\end{align}
where $E_{\kk \alpha } 
=  \sqrt{\left( \varepsilon_{\kk \alpha} - \mu \right)^2 + |\Delta_{\alpha}|^2}$, $\beta = 1 / k_\mathrm{B} T$, and $k_\mathrm{B}$ is Boltzmann's constant.
The fermion trace is given in the last line of the above equation. At zero temperature, we obtain 
\begin{eqnarray}
    F_0= \sum_{\alpha, \beta}\Delta_\alpha^\dagger V_{\alpha\beta}^{-1}\Delta_\beta + \sum_{\kk,\alpha} \left( \varepsilon_{\kk \alpha} - \mu \right)
    - \sum_{{\bf k},\alpha} E_{{\bf k}\alpha}, 
    \label{F_0_0}
\end{eqnarray}
where the last term is the zero-temperature limit of the fermion trace. Note that the fermion trace gives a negative contribution to $F_0$, which increases monotonically in magnitude as $|\Delta_\alpha|$ increases. Furthermore, the fermion trace is a convex function of $|\Delta_\alpha|$ at all $T$. This is particularly clearly seen at $T=0$, when it is given by $\sum_{{\bf k},\alpha} E_{{\bf k} \alpha}$. With the above assumptions, and using an energy cutoff $\omega_\alpha$ around each Fermi surface, the ${\bf k}$ summation in the last line may be performed analytically at $T=0$ to yield
\begin{eqnarray}
    F_0-F_{\mathrm{N}}  &=& \sum_{\alpha, \beta}\Delta_\alpha^\dagger V_{\alpha\beta}^{-1}\Delta_\beta 
    + \sum_\alpha N_\alpha \omega^2_\alpha ~
    G \left( \frac{|\Delta_\alpha|}{\omega_\alpha} \right) \label{F_0} \\
    F_{\mathrm{N}} &=&2 \sum_{{\kk},\alpha} \left( \varepsilon_{\kk \alpha} - \mu \right) \Theta(\mu - \varepsilon_{\kk \alpha}),
    \label{F_N}
\end{eqnarray}
where $G(x) = 1-\sqrt{1+x^2}-x^2 ~ \mathrm{arsinh}(1/x)$, 
$N_\alpha$ is a single-particle density of states at the Fermi surface of band $\alpha$, and $\Theta(x)$ is the Heaviside step function. $F_{\mathrm{N}}$ is the normal-state ground state energy, where the factor $2$ originates with a spin sum. Below, we will use the normal-state ground state energy as the reference zero point of $F_0$. 

\subsection{Interpretation of the gap equations}
In the above free energy $F(T)$, the $\{ \Delta_{\alpha} \}$ should be viewed as variational quantities that are to be determined by minimizing the free energy of the system.  A necessary condition for finding a minimum in $F(T)$ is that 
\begin{eqnarray}
    \frac{\partial F(T)}{\partial \Delta_\alpha} = 0 
    \label{eq:StationaryPoint}
\end{eqnarray}
for each band index $\alpha$.
The above stationary-point condition yields a set of coupled equations for the $N$ gaps $\Delta_\alpha$, the multiband gap equations. They read
\begin{eqnarray}
   \Delta_\alpha = \sum_{{\bf k},\beta}
   V_{\alpha \beta} \Delta_\beta ~ \chi_{{\bf k}\beta}(T), 
\label{eq:Gapequation}
\end{eqnarray}
where the Cooper-pair susceptibility is given by $\chi_{{\bf k}\beta}(T) = \tanh(\beta E_{{\bf k} \beta}/2)/2 E_{{\bf k}\beta}$. In general the gaps may be complex, and in the multiband case, the phases of the gaps in general do not cancel out of the gap equation. Equation\ \eqref{eq:Gapequation} may also be obtained directly from the definitions in Eq.\ \eqref{eq:Gaps} by computing $b_{{\bf k}\alpha}$ using the operators that diagonalize the Hamiltonian.
Equation \eqref{eq:Gapequation} has been employed as the starting point in numerous works on multiband $s$-wave superconductors \cite{Komendova2012, Wilson2013, Takahashi2014, Horhold2023, Benfenati2023}.

The standard procedure for determining which state is the thermodynamically stable state of the system is to select the solution of Eq.\ \eqref{eq:Gapequation} with the lowest free energy and compare this energy to the normal-state free energy. The ground state is then taken to be the one with the lowest free energy. Evaluating $F(T)$ at these two points is, however, not always sufficient to conclude that the system is superconducting, and we now highlight two further necessary conditions for $F(T)$ to be minimized. The first one is that $F(T)$ must attain a minimum at the solution in question, and the second one pertains to boundedness and will be considered in Sec. \ref{sec:diml_reduction_general}. In order to have a minimum in $F(T)$, the Hessian, whose elements are
\begin{eqnarray}
   \mathcal{H}_{ij} \equiv \frac{\partial^2 F(T)}{\partial x_i \partial x_j},
\label{eq:PositiveDefiniteHessian}
\end{eqnarray}
should be positive definite when computed at the stationary point. Here, the set $\{x_i\}$ contains the $2N-1$ independent variables $\{ |\Delta_\alpha|,\theta_{\beta \gamma} \}$. There are $N$ gap amplitudes and $N - 1$ independent phase differences $\theta_{\beta \gamma} \equiv \theta_\beta - \theta_\gamma$, where $\theta_\beta$ is defined by $\Delta_\beta = |\Delta_\beta|\mathrm{e}^{i\theta_\beta}$. 
Both conditions Eq.\ \eqref{eq:StationaryPoint} and the positive definiteness of $\mathcal{H}$ need to be satisfied for any nontrivial solution to Eq.\ \eqref{eq:Gapequation} to be physically meaningful in the sense of corresponding to {\it thermodynamically stable} superconducting states. As mentioned, even if both conditions are satisfied, it is still not a {\it sufficient} requirement for a solution to correspond to a stable superconducting state, since the minimum could be a local minimum. Such a local minimum could correspond to a metastable superconducting state, whereas a stable superconducting state would have to be a solution corresponding to a global minimum. On the other hand, as we shall see in the following sections, nontrivial solutions to Eq.\ \eqref{eq:Gapequation} could very well exist that satisfy Eq.\ \eqref{eq:StationaryPoint} {\it without} having a positive definite $\mathcal{H}$. The solutions then could, and in fact very often do, correspond to a maximum or a saddle point in $F(T)$, both of which are unphysical if \emph{all} the $x_i$'s are to be regarded as independent variational parameters. As we are about to see, however, there are situations in which constraints must be put on $\{x_i\}$ before minimizing $F(T)$.

The simplest system that necessitates constraints is one where $V$ (and hence $V^{-1}$) is a diagonal matrix. Such a system is a decoupled set of one-band superconductors with nontrivial physically meaningful solutions $\Delta_{\alpha}$ to the decoupled gap equations {\it only} for those diagonal elements that satisfy $V_{\alpha \alpha } > 0$.
The remaining gaps, those with repulsive $V_{\alpha \alpha } < 0$, are put to zero. This procedure serves as motivation for the dimensionality reduction outlined in Sec. \ref{sec:diml_reduction_general}, 
which is a generalization to 
systems with interband interactions, $V_{\alpha \beta} \neq 0; \alpha \neq \beta$. Then, the situation is less clear, and care has to be taken to verify that nontrivial solutions to Eq.\ \eqref{eq:Gapequation} actually represent physical superconducting states.

\subsection{Dimensional reduction of the order-parameter space}
\label{sec:diml_reduction_general}
\begin{figure*}[t]
    \centering    \includegraphics[width=\textwidth,keepaspectratio]{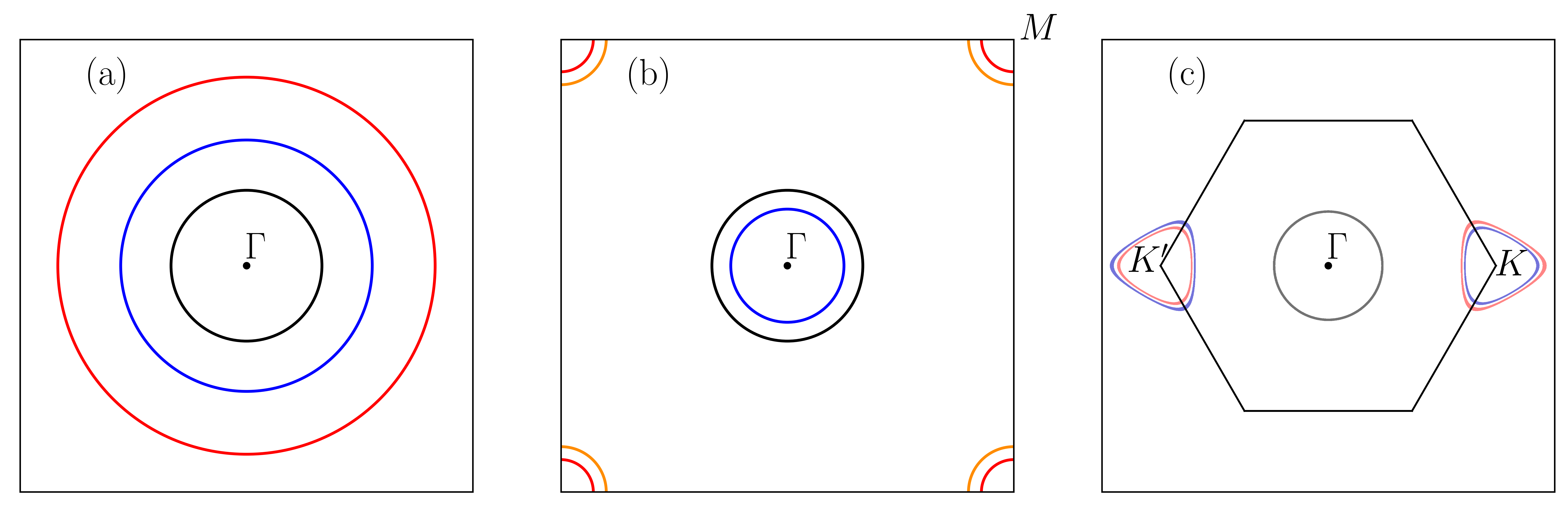}
    \caption{Schematic figure of three different examples of multisheeted Fermi surfaces centered around high-symmetry points. (a) The Fermi surfaces of a simple multiband model with all bands centered around the $\Gamma$ point. (b) The Fermi surface with two bands located at the zone center and two bands centered in the zone corners around $M$. Such a model for the Fermi surface is relevant in simplified descriptions of the Fermi surfaces in iron pnictide superconductors such as \ch{FeAs}. (c) One Fermi pocket centered around $\Gamma$ and two Fermi pockets located around $K$ and $K^{\prime}$ points of the Brillouin zone (enclosed by the black line) of a honeycomb or triangular lattice. Strong Ising spin-orbit coupling lifts the spin degeneracy near the $K/K^{\prime}$ points, causing the two spin-polarized bands, marked in red and blue, to be inverted in going from $K$ to $K^{\prime}$. Such a multisheeted Fermi surface is relevant to the transition metal chalcogenide \ch{NbSe_2}. The model in Eq.\ \eqref{eq:NbandBCS} is applicable to all three cases.}
    \label{fig:fermi_surfaces}
\end{figure*}

The abovementioned stationary points are crucially dependent on the  term 
$\sum_{\alpha \beta} \Delta^{\dagger}_\alpha V^{-1}_{\alpha\beta} \Delta_\beta$ 
in Eqs. \eqref{F_T} and \eqref{F_0_0}, which brings us to the second condition on the free energy that must be met, namely boundedness from below. We note that this term will also govern the large-$|\Delta|$ behavior of $F_0$ since it scales quadratically with $\Delta$ whereas the fermion trace is negative and asymptotically linear in $\Delta$. 
This can be seen more clearly by performing a change of basis, rewriting $\sum_{\alpha \beta} \Delta^{\dagger}_\alpha V^{-1}_{\alpha\beta} \Delta_\beta$ as $\sum_{\gamma=1}^N \lambda_{\gamma} |\Psi_{\gamma}|^2$,
where $\Psi_{\gamma} = \sum_{\beta} S_{\gamma \beta}^{-1} \Delta_\beta$ and $S_{\alpha \beta}$ are matrix elements of the modal matrix of $V^{-1}$, $S^{-1} V^{-1} S = \rm{Diag}(\lambda_1,\dots,\lambda_N)$. Here, we use the convention that eigenvalues $\lambda_1,\dots,\lambda_N $ are ordered in descending order. If $V$ and $V^{-1}$ violate the criterion of being positive definite, then we have $\lambda_{\gamma} \leq 0, \gamma \in \{M+1,\dots,N\}$, while $\lambda_{\gamma} > 0, \gamma \in \{1,\dots,M\}$ for some $0 \leq M < N$.
Thus, an increase of the rotated order parameter along the direction of a $\Psi_\gamma$ with $\gamma>M$ will always lower the free energy, seemingly leaving the free energy \emph{unbounded from below} and the mean-field theory ill-defined.
The correct way to proceed is then to consider a mean-field theory, not in terms of the original $N$ components $\Delta_\alpha$,  but in terms of the $M$ components $\Psi_\gamma$ that belong to positive eigenvalues. All components $\{ \Delta_{\alpha}\}$ are retained, but in addition to Eq.\ \eqref{eq:Gapequation}, there are $N-M$ constraints on these amplitudes. This reduces the dimensionality of the multicomponent superconducting order-parameter space. The constraints are given by 
\begin{eqnarray}
\sum_{\beta} S^{-1}_{\gamma \beta } \Delta_{\beta} = 0, \gamma \in \{M+1,\dots,N\}.
\label{eq:constraints}
\end{eqnarray}

For diagonal $V$, $S^{-1}_{\gamma \beta } = \delta_{\gamma \beta}$ and Eq.\ \eqref{eq:constraints} reduces to the obvious $\Delta_{\gamma} = 0, \gamma \in \{M+1,\dots,N\}$, consistent with the case of diagonal $V$ considered previously. With nondiagonal elements of $V$ present, the situation is more complicated, but the principle is nevertheless the same: There will be a reduced number of nonzero components of the order parameter, $\{ \Psi_{\gamma} \}, \gamma \in \{1,\dots,N-M\}$. These $\{ \Psi_{\gamma} \}$ could involve a linear combination of all $\{ \Delta_{\alpha}\}$, but there will only be $N-M$ such linear combinations. Barring degeneracies in $\lambda_{\gamma}$, these linear combinations are all orthogonal. In the presence of degenerate $\lambda_{\gamma}$, the degenerate subspace of linear combinations can be re-orthogonalized by a Gram-Schmidt procedure. 
Working in the original space of order-parameter components, $\{ \Delta_{\alpha} \}, \alpha \in \{1,\dots,N\}$ with no further constraints could lead to solutions of Eq.\ \eqref{eq:Gapequation} corresponding to maxima, saddle points, or local minima in the free energy.
Moreover, in such cases, the mean-field theory would be ill-defined as there is no lower bound on its free energy, meaning that any higher-order expansion around the stationary points will be futile since fluctuations will eventually drive the system away from these. Imposing constraints (if needed) on fluctuations in the order parameter may thus also be important, especially when using effective field theories \cite{Carlstrom2011, Vagov2012, Speight2019, Speight2023} to calculate physical quantities in multiband systems. 
Dimensional reduction of the the order-parameter space for a singular interaction matrix $V$ has previously been noted \cite{Benfenati2023}. Equation \eqref{eq:constraints} extends the reduction to also encompass interaction matrices of arbitrary size that are not positive definite. Similar results have previously been discussed for the case $N=3$, but only in the absence of intraband interactions \cite{Garaud2017}.

From the arguments above, we can also extend the original criterion for one-band BCS $s$-wave superconductivity, namely that the interaction must be attractive, to multiband systems. For systems with momentum-independent effective electron interactions, the unconstrained superconducting free energy is bounded from below if and only if $V^{-1}$ is positive definite. If not, care must be taken to reduce the dimensionality of the superconducting order-parameter space, 
such that the thermodynamically stable state can be identified.
We note that $V$ and $V^{-1}$ have the same eigenvectors, and their eigenvalues are reciprocals. So in the following, we will mainly focus on $V$ since it offers a more transparent connection to the underlying scattering processes. For similar reasons, only the case of zero-temperature systems will be considered, allowing the use of the analytical expression for the free energy in Eq.\ \eqref{F_0}. This is not an essential simplification as the discussion presented in this section is straightforwardly generalized to nonzero temperatures, which can be seen from the fact that the Fermion-trace term gives a strictly negative contribution to the free energy, depends only on $\{ |\Delta_{\alpha}| \}$ and is a convex function of these moduli for all values of $T$.

In the next sections, we consider the two-band case in detail and also comment on the case of three, four, and five bands. The two-band case has the virtue of being quite transparent. A main conclusion is that the mechanism that constrains thermodynamically stable superconductivity by increasing interband attraction or repulsion appears to be operative for any number of bands, where the situation rapidly becomes more complicated. Many of our comments are pertinent to recent theoretical works on such systems, for instance on the multiband superconductor \ch{NbSe_2}. 

\section{From one to two bands}\label{sec:2_bands}
The point of this section is to substantiate our claims in the previous section. We will illustrate in detail that the minimization process for a conventional one-band superconductor, as considered in the original BCS paper, Ref.\ \cite{Bardeen1957}, cannot be straightforwardly extended to a multiband superconductor, using the two-band system as an example.
 
First, we consider a conventional one-band superconductor in the zero-temperature limit. From Eq.\ \eqref{F_0}, it follows that the free energy only depends on the modulus of the gap. To distinguish the one-band case from the multiband case where both the modulus and phase of the gaps determine the free energy, we denote the modulus of the one-band superconducting gap as $\tilde{\Delta}\geq 0$. The derivative of the one-band free energy, denoted $F_0^1$, is 
\begin{equation}
    \frac{\partial F_0^1}{\partial\tilde{\Delta}} = 2\tilde{\Delta} N_1\bigg[ \frac{1}{N_1V_{11}} -\mathrm{arsinh}\left(\frac{\omega_1}{\tilde{\Delta}} \right) \bigg].
    \label{first_derivative_F_0}
\end{equation}
Extremizing $F_0^1$ yields two solutions, namely the trivial solution $\tilde{\Delta} = 0$ and the nontrivial BCS solution $\tilde{\Delta}_{\mathrm{BCS}} \equiv \omega_1/\sinh(\lambda_1^{-1})$ with $\lambda_1 = N_1V_{11}$. In order to determine the preferred state of the system, the second derivative at each of the extrema must be evaluated. At $\tilde{\Delta} = 0$, the second derivative diverges logarithmically to minus infinity, whereas the second derivative at the BCS solution is always positive,
\begin{align}
    \frac{\partial^2 F_0^1}{\partial\tilde{\Delta}^2}\bigg\rvert_{\tilde{\Delta} = \tilde{\Delta}_{\mathrm{BCS}}} &= 2N_1\frac{\omega_1}{\sqrt{\omega_1^2 + \tilde{\Delta}^2_{\mathrm{BCS}}}}.
    \label{second_derivative_BCS}
\end{align}
Thus, one concludes that for a one-band system with an attractive interaction at zero temperature, the energy of the superconducting state is lower than the normal state, regardless of the strength of the attraction \cite{Bardeen1957}. Furthermore, we note that because the free energy has a global energy minimum at $\tilde{\Delta} = \tilde{\Delta}_{\mathrm{BCS}}$, it is bounded from below. If the interaction is repulsive, the free energy is unbounded from below, and only the trivial solution to the gap equation exists, as expected since a repulsive interaction would render the system a Fermi liquid. These comments also apply to the multiband case with diagonal interaction matrix elements. In the following, we will highlight that in the multiband case with interband interactions, one might find nontrivial solutions to the stationary-point condition, for which the stationary point is not a minimum in the free energy.
\subsection{Stability analysis of the unconstrained two-band case}
\begin{figure*}[t]
    \centering
    \includegraphics[width=\textwidth,keepaspectratio]{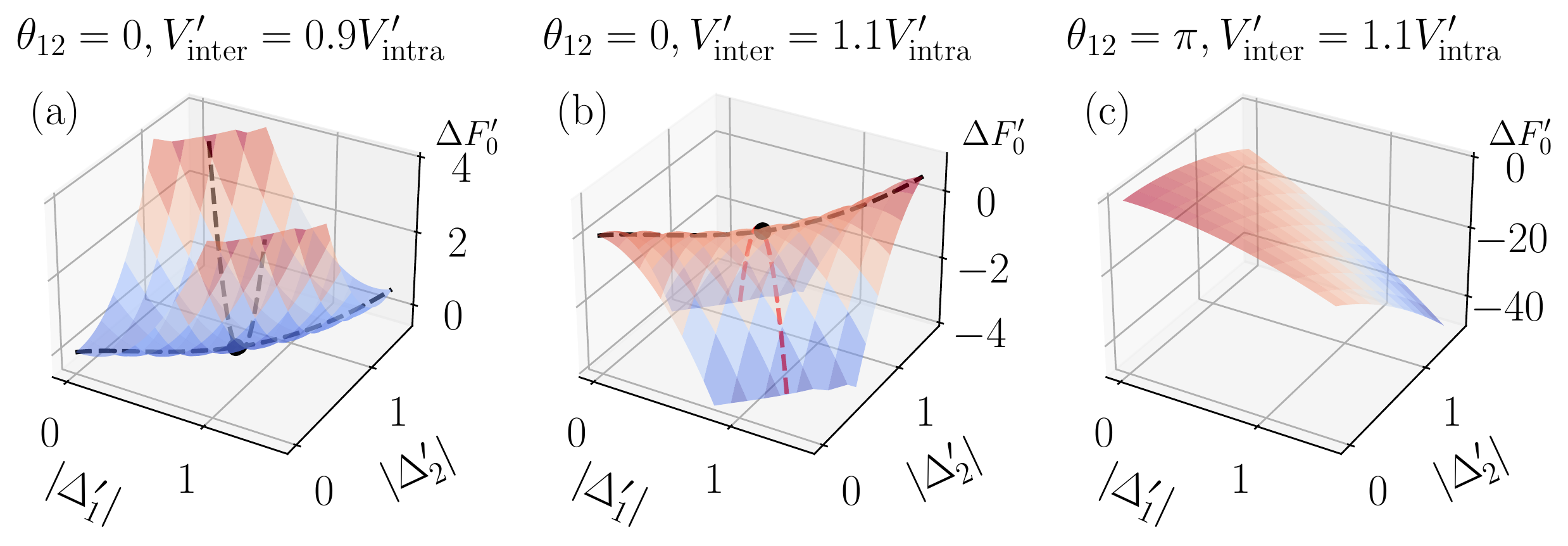}
    \caption{The dimensionless free-energy difference $\Delta F'= (F_0 - F_{\mathrm{N}})/(N_{\mathrm{F}} \omega_c^2)$ at zero temperature of a two-band superconducting system in mean-field theory as a function of the dimensionless gap amplitudes $|\Delta'_\alpha|= |\Delta_\alpha|/\omega_c$. The bands have equal density of states at the Fermi surface and energy cutoff, denoted $N_{\mathrm{F}}$ and $\omega_c$, respectively.
    The phase difference between the gaps $\theta_{12}$ varies across the three panels. Dimensionless interactions $V'_{\alpha\beta} = V_{\alpha\beta}N_{\mathrm{F}}$ are used. All interactions are attractive, and the largest interaction in each panel is $0.5$. The free-energy surface is colored red (blue) when it is large (small). The black dot in panels (a) and (b) marks a stationary point in the free energy, and the eigenvectors of the associated Hessian matrix belonging to positive eigenvalues are marked by dashed black lines. In panel (b), there is only one black line since only one of the eigenvalues is positive. The other eigenvector is marked by the dashed red line.}
    \label{fig:energy_landscape}
\end{figure*}
When considering multiband systems, it is particularly instructive to consider the case of $N=2$ since both $\lambda_\alpha$ and $\Psi_\alpha$ can be expressed directly in a transparent manner by using the matrix elements of $V$. This two-band system was first considered in seminal works on multiband superconductivity in transition metals with significant scattering between $s$ and $d$ orbitals (ignoring the possibility of complex-valued $\{\Delta_{\alpha} \}$)\cite{Suhl1959, Kondo1963}. In keeping with their notation, we denote the general two-band scattering matrix $V$ as
\begin{equation}
    V = \begin{pmatrix}
        V_{ss} & V_{sd} \\
        V_{sd} & V_{dd}
    \end{pmatrix},
    \label{V_def}
\end{equation}
where $V_{ss}$ and $V_{dd}$ describe intraband interaction strength, and $V_{sd}$ is the interband interaction strength. To find the eigenvalues of $V^{-1}$, we find the eigenvalues of $V$, as their reciprocals are eigenvalues of $V^{-1}$. The eigenvalues of $V$ are
\begin{align}
    \lambda_\pm = \frac{1}{2}\bigg[ V_{ss} + V_{dd} \pm \sqrt{4V_{sd}^2 + (V_{ss} - V_{dd})^2}\bigg],
    \label{eigvals_two_band}
\end{align}
and accordingly, the eigenvalues of $V^{-1}$ are $\lambda^{-1}_\pm$. Imposing $\lambda_\pm >0$ is equivalent to imposing the constraints 
    $V_{ss} + V_{dd}>0,
V_{ss}V_{dd}>V_{sd}^2$. 
It follows that in order for the interaction matrix to be positive definite and the free energy to be bounded from below in the full $(\Delta_1, \Delta_2)$ plane, both intraband interactions must be attractive and stronger than the interband interaction.
In such a case, we may search for a nontrivial solution to Eqs. \eqref{eq:StationaryPoint} and \eqref{eq:Gapequation} in an otherwise unconstrained order-parameter space $(\Delta_1, \Delta_2 )$. Imposing $\lambda_{-} < 0$ and $\lambda_+ > 0$ corresponds to $V^2_{sd} > V_{ss} V_{dd}$. This holds for strong enough interband interactions if both intraband interactions are attractive or both are repulsive, or for any interband interaction if one of the intraband interactions is repulsive and the other is attractive. The stationary point of $F$ must then be sought with one additional constraint, Eq.\ \eqref{eq:constraints}. Finally, if both intraband interactions are repulsive, and $V_{ss} V_{dd} > V^2_{sd}$, then $\lambda_{\pm} < 0$. There are then two additional constraints Eq.\ \eqref{eq:constraints}, rendering $(\Delta_1,\Delta_2)=0$.

To further elucidate this, we consider the case of equal attractive intraband interactions, denoted $V_{ss} = V_{dd} = V_{\mathrm{intra}}>0$ for $V_{\mathrm{intra}}$ both smaller and larger than $V_{\mathrm{inter}}$, where $V_{\mathrm{inter}}$ denotes the interband interaction strength.
Furthermore, for simplicity we assume the energy cutoff and density of states at the Fermi surface, denoted $\omega_c$ and $N_{\mathrm{F}}$, respectively, to also be equal.
Component $\alpha$ of the superconducting order parameter is a complex number with phase $\theta_\alpha$, so $\Delta_\alpha \Delta^\dagger_\beta =|\Delta_\alpha| |\Delta_\beta| \exp(i \theta_{\alpha \beta})$, where $\theta_{\alpha \beta} = \theta_\alpha - \theta_\beta$. 
The phase difference between the two bands, denoted $\theta_{12}$, only enters in the free energy as part of the term $2V_{12}^{-1}|\Delta_1||\Delta_2|\cos(\theta_{12})$. It is thus determined by the equation
\begin{equation}
\cos(\theta_{12}) = -\mathrm{sgn}(V_{12}^{-1}) = \mathrm{sgn}(V_{\mathrm{inter}}/\mathrm{det}(V)),
\label{phase_diff_selection_rule}
\end{equation}
causing a $0$-$\pi$ transition \cite{Iskin2006} as $V_{12}^{-1}$ switches sign. 
%The degree of freedom offered by $\theta_{12}$ renders the sign of $V_{12}^{-1}$ irrelevant.
To determine if the superconducting state is energetically favorable, we define $\Delta F = F(T) - F_{\mathrm{N}}$ where $F_{\mathrm{N}}$ is the normal-state free energy. At zero temperature, the analytical expression for $F_0$ given in Eq.\ \eqref{F_0} replaces $F(0)$ and will be used unless stated otherwise.
Figure \ref{fig:energy_landscape} shows the dimensionless free energy difference $ \Delta F'\equiv \Delta F/(N_{\mathrm{F}} \omega_c^2)$ as a function of the dimensionless variables $|\Delta'_\alpha| = |\Delta_\alpha|/\omega_c$ for three different systems. In Fig.\ \ref{fig:energy_landscape} (a) we consider a system satisfying $\lambda_{\pm} >0$, whereas this does not hold for the systems in Fig.\ \ref{fig:energy_landscape} (b, c). The phase difference minimizing the energy, $\theta_{12}=0$, is chosen in Fig.\ \ref{fig:energy_landscape} (a), as $V_{12}^{-1}<0$. The location of the global minimum can be obtained by solving the gap equation in Eq.\ \eqref{eq:Gapequation} analytically and is equal to
\begin{equation}
    |\Delta_{\mathrm{extremum}}| = \omega_c/\sinh[(V_{11}^{-1} + |V_{12}^{-1}|)/N_{\mathrm{F}}]
    \label{solution_to_simple_two_bands}
\end{equation}
for both gaps. Note that the free energy of this system would still be bounded from below if $\theta_{12}=\pi$, but the location of the energy minimum would change. When $V_{\mathrm{inter}}>V_{\mathrm{intra}}$, the free energy is unbounded from below as illustrated in Figs.\ \ref{fig:energy_landscape} (b, c). For $\theta_{12} =0$ in Fig.\ \ref{fig:energy_landscape} (b), there exists a saddle point in the free energy, marked by a black dot, again situated at the point where the modulus of both gaps are equal to $|\Delta_{\mathrm{extremum}}|$. Since the gradient of the free energy is zero at this point, it is a nontrivial solution to the gap equation. It is also at lower energy than the normal state. The standard procedure for determining the ground state would therefore select this as the ground state even though it is thermodynamically unstable when the constraints in Eq.\ \eqref{eq:constraints} are not enforced. The existence of this solution relies on choosing $\theta_{12}$ opposite to that of Eq.\ \eqref{phase_diff_selection_rule}, i.e., by choosing the phase difference maximizing the free energy. When choosing the phase difference in accordance with Eq.\ \eqref{phase_diff_selection_rule}, $\theta_{12}=\pi$, the only solution to the gap equation is the trivial one, and the free energy is lower in Fig.\ \ref{fig:energy_landscape} (c) than (b) for all finite $\boldsymbol{\Delta}$. We also note that the results presented for the simple systems in Fig.\ \ref{fig:energy_landscape} straightforwardly generalize to two-band systems where $V_{ss} \neq V_{dd}$ and $N_s\neq N_d$. The main difference is that Eq.\ \eqref{solution_to_simple_two_bands} no longer describes nontrivial solutions to the gap equation, which must be solved numerically. The qualitative behavior of the free energy remains the same, also if the strength of the interactions is increased or decreased.

\subsection{Dimensional reduction of the order-parameter space in the two-band case}
The associated eigenvectors of $\lambda_\pm$ are $1/\sqrt{2}\begin{pmatrix}
    1 & \pm 1
\end{pmatrix}^{\mathrm{T}}$. So by enforcing $\Psi_-=0$ and using the expressions for the eigenvectors in the modal matrix $S$, one obtains the constraint
\begin{equation}
    -\Delta_1 + \Delta_2 = 0
    \label{simple_constraint}
\end{equation}
from Eq.\ \eqref{eq:constraints}. Equation \eqref{simple_constraint} can only be fulfilled when $|\Delta_1|=|\Delta_2|$ and $\theta_{12} = 0$. However, since $V_{12}^{-1}>0$, and thus, by Eq.\ \eqref{phase_diff_selection_rule}, the most energetically favorable value of $\theta_{12}$ is $\pi$.
This seeming contradiction is simply a manifestation of an earlier statement, namely that one may search in an unconstrained order-parameter space only when $V$ is positive definite. If not, there are constraints that must be satisfied, also when they prohibit following the selection rule in Eq.\ \eqref{phase_diff_selection_rule}.
Moreover, by enforcing the hard constraint in Eq.\ \eqref{simple_constraint}, we can see that this is equivalent to confining the order-parameter space to the dashed black line along the spine of the free-energy landscape in Fig.\ \ref{fig:energy_landscape} (b).
We stress that the Hessian matrix in Eq.\ \eqref{eq:PositiveDefiniteHessian} evaluated at the saddle point becomes positive definite only when applying the constraint. In Fig.\ \ref{fig:energy_landscape}, this is illustrated by having dashed black (red) lines in the direction of eigenvectors of the Hessian belonging to positive (negative) eigenvalues. Since the constraint in Eq.\ \eqref{simple_constraint} prohibits fluctuations in any direction but $|\Delta_1|= |\Delta_2|$, the Hessian is positive definite at the stationary point $|\Delta_1|= |\Delta_2| = \Delta_{\mathrm{extremum}}$. The stationary point is a global energy minimum.

\begin{figure}
    \centering
    \includegraphics[width=\linewidth,keepaspectratio]{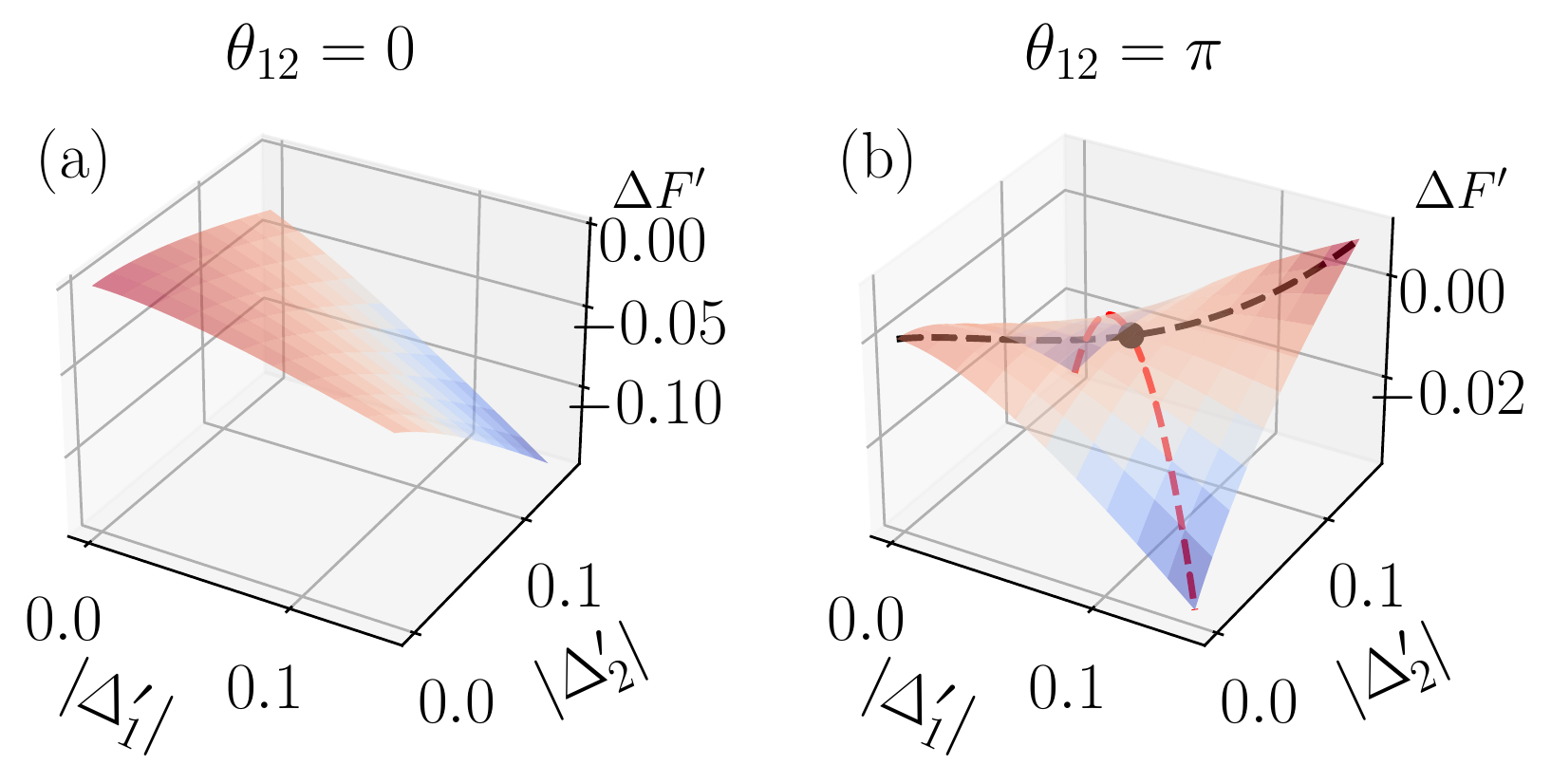}
    \caption{The same as in Fig.\ \ref{fig:energy_landscape}, but with repulsive interactions $V_{\alpha\beta}<0$. In both panels, the interaction strengths are $V'_{\mathrm{inter}} = -1.3$ and $V'_{\mathrm{intra}} = -1.0$. 
    }
    \label{fig:repulsive}
\end{figure}

The free energy of a system with repulsive interactions $V_{\alpha\beta}<0$ and $|V_{ss} V_{dd}|<V_{sd}^2$ is illustrated in Fig.\ \ref{fig:repulsive} (a,b), with $\theta_{12} =0$ and $\theta_{12} =\pi$, respectively, values that both extremize the free energy.
For simplicity, we again look at a system with equal intraband interactions $V_{ss}=V_{dd}=-|V_{\mathrm{intra}}|$ and equal density of states at the Fermi surface. In both figures, there is a stationary point in the free energy at $(|\Delta_\alpha| = 0, |\Delta_\beta| = 0)$. A saddle point is also present in Fig.\ \ref{fig:repulsive} (b), marked by a black dot.
Since $|V_{ss} V_{dd}|<V_{sd}^2$, $\lambda_+\lambda_-<0$ as in the systems in Fig.\ \ref{fig:energy_landscape} (b, c). However the constraint in Eq.\ \eqref{eq:constraints} now becomes
\begin{equation}
    \Delta_1 = -\Delta_2.
    \label{simple_constraint_repulsive}
\end{equation}
So although the choice $\theta_{12}=0$ renders the term $\sum_{\alpha \beta} \Delta^{\dagger}_{\alpha} V^{-1}_{\alpha \beta} \Delta_{\beta}$ lower than the choice $\theta_{12} = \pi$ (since $V_{12}^{-1}<0$), it violates the constraint akin to the two-band system considered previously. Choosing $\theta_{12}=\pi$ is thus necessary to satisfy the order-parameter constraint. The free energy of the system with $\theta_{12}=\pi$ is shown in Fig.\ \ref{fig:repulsive} (b).
Similar to the case of Fig.\ \ref{fig:energy_landscape} (b), along the line $|\Delta_1|=|\Delta_2|$, marked by a dashed line, there is a global minimum in the free energy located at $|\Delta_1|= |\Delta_2| = \Delta_{\mathrm{extremum}}$.
A common criterion for having $s$-wave superconductivity in (partly) repulsive BCS systems found in the literature is $|V_{ss} V_{dd}|<V_{sd}^2$ \cite{Roldan2013, Horhold2023} which has been motivated mainly by requiring the existence of a nontrivial solution to the gap equation. Now it can be further understood as a requirement for one of the eigenvalues to turn positive such that the emergent saddle point in the unconstrained free energy becomes a global energy minimum when applying the constraint in Eq.\ \eqref{simple_constraint_repulsive}. 

The case of a system with only repulsive interactions is especially relevant in the context of superconducting iron pnictides \cite{Stewart2011}, as repulsive interactions are believed to play an essential part in driving their superconducting transition \cite{Chubukov2009}. Depending on the pnictide, there will be several hole and/or electron pockets, which can be illustrated in a simplified manner as in Fig.\ \ref{fig:fermi_surfaces} (c), leaving Eq.\ \eqref{eq:NbandBCS} as a natural starting point for $s$-wave superconductivity since there are multiple disjoint Fermi surfaces. While the two-band model considered here is simpler than the four-pocket model in Fig.\ \ref{fig:fermi_surfaces} (c), it serves as an indication that some constraints must be put on the order parameter in more realistic models, as it is unlikely that all eigenvalues of a repulsive interaction matrix are positive. These constraints will be more complicated than that of Eq.\ \eqref{simple_constraint_repulsive}, but they are easily extracted numerically using Eq.\ \eqref{eq:constraints}. We leave this for future investigations.

We close this section by noting that the constraint of two-band systems with $\lambda_+\lambda_-<0$ (equivalently $V^2_{12} > V_{11} V_{22}$) can be derived for a general interaction matrix. From Eq.\ \eqref{eq:constraints} it reads
\begin{eqnarray}
-V_{12} \mathrm{e}^{-i \theta_{12}} |\Delta_1| + (d_1 + d_2) |\Delta_2| = 0,
\label{eq:general_two_band_constraint}
\end{eqnarray}
where $d_1 = (V_{11} - V_{22})/2$, and $d_2 = \sqrt{(V_{12})^2 + d_1^2}$. Thus, a constraint on the phase $\theta_{12}$ is given by $\theta_{12} = 0,\pi$, and $V_{12} \cos(\theta_{12}) > 0$. Hence, $|V_{12}| |\Delta_1| = (d_1+d_2) |\Delta_2|$, which defines the line in $(|\Delta_1|,|\Delta_2|)$ space along which a minimum in the free energy must be sought. By Eq.\ \eqref{eq:general_two_band_constraint}, one immediately obtains the constraints in Eqs.\ \eqref{simple_constraint} and \eqref{simple_constraint_repulsive}, which only differ by a minus sign, owing to the fact that $\mathrm{sgn}(V_{12})$ is different for the two systems.

\section{Strong interband interactions in multiband systems}\label{sec:TMD}
In the previous section, we focused on the sign of the eigenvalues of the interaction matrix $V$ for two-band systems and the implications nonpositive eigenvalues have for constraining the components $\{ \Delta_{\alpha}\}$ of the superconducting order parameter. Analytical expressions for eigenvalues of general symmetric $N \times N$ matrices rapidly become intractable for $N>2$. It is difficult to gain insight into the properties of such systems compared to the case of $N=2$. However, in this section, we will show that the connection between the number of nonpositive eigenvalues and the reduction of the order-parameter space derived for two-band systems still serves as useful guidance for $N$-band systems. 

In the monolayer TMDs \ch{NbSe_2} and \ch{TaS_2}, the Fermi surface may be illustrated as in Fig.\ \ref{fig:fermi_surfaces} (c) \cite{DeLaBarrera2018}. 
These systems are superconductors \cite{Xi2016, Navarro-Moratalla2016} with three disjoint Fermi surfaces. They can thus be treated with the multi-patch method \cite{Perali2000, Perali2001}, and are well-described by the Hamiltonian in Eq.\ \eqref{eq:N-band_MFT_BCS} where each $\Delta_\alpha$ is isotropic within a specific Fermi pocket. 
This remains the case even when including strong Ising spin-orbit coupling in the Fermi pockets centered around $K$ and $K'$ such that $\varepsilon^\uparrow(\kk) = \varepsilon^\downarrow(-\kk)$ \cite{Wickramaratne2020, Horhold2023}, as illustrated in Fig.\ \ref{fig:fermi_surfaces} (c).
Scanning tunneling spectroscopy has been used to experimentally probe the multigap nature of both bulk $2H$-\ch{NbSe_2} and few-layer (monolayer) \ch{NbSe_2}. Tunneling conductance results for bulk \ch{NbSe_2} seem to support the presence of two-band superconductivity \cite{Dvir2018}, with the possibility of having an additional contribution from a third gap \cite{Noat2015}. 
A multipocket model can also reproduce experimental data in few-layer systems. However, this has been argued to be a less likely explanation \cite{Kuzmanovi2022}, especially when there is disorder in the sample. In addition to the rich physics of monolayer TMDs, they also often exhibit strong interactions between the $K$ and $K'$ pockets. For our purposes, they serve as an instructive and relevant example for demonstrating how strong interband interactions influence superconductivity. 

Even though the Fermi pockets in \ch{NbSe_2} originate with the same band, we use the terms band and pocket interchangeably, as each pocket may have a unique dispersion relation at the Fermi surface. Due to the momentum restrictions BCS pairing imposes, the scattering processes near the Fermi surface will depend on which pockets the interacting electrons belong to. For intraband interactions $V_{\alpha\alpha}$, the average magnitude of the momentum transfer $|\qq| \equiv|\kk -\kk'|$, will be different for the pockets centered around $K$ and $K'$ than for the pocket centered around $\Gamma$. Similarly, the interband interactions between the $\Gamma$ band and the $K/K'$ bands, denoted $V_{\Gamma K}=V_{\Gamma K'}$, will on average have slightly larger $|\qq|$ than $V_{\alpha\alpha}$, whereas interband interactions between $K$ and $K'$ pockets $V_{KK'}$ will have the largest momentum transfer. Hence $V_{KK}$ and $V_{\Gamma\Gamma}$ ($V_{KK'}$) correspond to the long (short)-wavelength part of the interaction. If all interactions are only weakly dependent on $\kk$ and $\kp$, its $\qq$ dependence can be averaged out such that each of the possible intra-/interband interactions in the TMD may be replaced by their average values \cite{Roldan2013, Horhold2023}. The system is then described by the $s$-wave Hamiltonian in Eq.\ \eqref{eq:NbandBCS} with an effective interaction matrix given by
\begin{equation}
    V_{\mathrm{TMD}} = \begin{pmatrix}
        V_{KK} & V_{KK'} & V_{\Gamma K} \\
        V_{KK'} & V_{KK} & V_{\Gamma K} \\
        V_{\Gamma K} & V_{\Gamma K} & V_{\Gamma \Gamma} 
    \end{pmatrix}.
    \label{V_TMD_def}
\end{equation}
$V_{\mathrm{TMD}}$ is a relatively simple symmetric matrix and describes the many types of interactions well, as long as they can be averaged in a meaningful way.

Consider the case where the free energy is assumed to be bounded from below in an unrestricted order-parameter space. This requires a positive definite $V_{\mathrm{TMD}}$. An alternative criterion for a symmetric matrix to be positive definite is that all leading principal minors must be positive. Thus, $V_{\mathrm{TMD}}$ is positive definite if and only if $V_{KK}>0$, $V_{KK}^2>V_{KK'}^2$ and $V_{KK}^2V_{\Gamma\Gamma}>V_{\Gamma\Gamma}V_{KK'}^2 + 2V_{K\Gamma}^2(V_{KK} - V_{KK'})$, see Appendix \ref{app:interaction_matrices}.
The first two criteria are the same as in the two-band case.
By assuming these to be fulfilled, one finds that this further imposes $V_{\Gamma\Gamma}>0$, such that the last criteria can be written as 
\begin{equation}
    V_{KK}^2>V_{KK'}^2 + 2\frac{V_{K\Gamma}^2}{V_{\Gamma\Gamma}}(V_{KK} - V_{KK'}).
    \label{V_TMD_inter_stability}
\end{equation}
The role of the interband interactions is evident in Eq.\ \eqref{V_TMD_inter_stability}. Increasing either $V_{KK'}$ or $V_{K\Gamma}$ leads to an increase in the right-hand side, thus eventually violating the inequality. If this is the case, ensuring a lower bound on the free energy requires the constraint
Eq.\ \eqref{eq:constraints} on the order-parameter components to be imposed. Increasing intraband interactions has the opposite effect.

$V_{\mathrm{TMD}}$ can describe different types of interactions in TMDs, but the eigenvalue spectrum of $V_{\mathrm{TMD}}$ will depend on the numerical values of $V_{\alpha\beta}$ for the specific TMD under consideration. Thus, exactly which interaction mechanisms one includes when calculating the matrix elements of $V_{\mathrm{TMD}}$ are important. In the case of Coulomb repulsion, $V_{KK}$ and $V_{KK'}$ describe interactions from the tail of the screened Coulomb interaction and the short-ranged part \cite{Roldan2013, Horhold2023}, respectively, such that $|V_{KK'}|>|V_{KK}|$. For attractive interactions mediated by phonons, the ratio between $V_{KK'}$ and $V_{KK}$ is not as transparent since it requires detailed calculations of how the phonons of the system couple to the electrons. The electron-phonon coupling for a material can be calculated using a tight-binding model \cite{Thingstad2020} or from first-principles calculations \cite{Kaasbjerg2012, Zheng2019}.
In \ch{NbSe_2}, the largest contribution to the effective attractive interaction is mediated by short-wavelength phonons, especially if the system is assumed not to coexist with a charge-density wave \cite{Zheng2019, Das2023}. Using the $\qq$-averaged model, this corresponds to $V_{KK'}$ being the largest matrix element in $V_{\mathrm{TMD}}$, causing a violation of one or more of the inequalities above. 
However, by including other interactions, $V_{\alpha \beta}$ may change and cause $V_{\mathrm{TMD}}$ to become positive definite. For instance, including Coulomb repulsion will affect $V_{KK'}$ more than $V_{KK}$, which could, in turn, cause $V_{KK}$ to be larger than $V_{KK'}$. Spin fluctuations have also been observed to play an influential role in \ch{NbSe_2} \cite{Wickramaratne2020,Das2023} and could similarly affect the spectrum of $V_{\mathrm{TMD}}$.
The point here is not to make any quantitative statements about $V_{\mathrm{TMD}}$, but rather to bring attention to the pivotal role of scattering between the Fermi pockets in determining its spectrum.
Thus, TMDs with strong interband interactions are subject to the constraints in Eq.\ \eqref{eq:constraints}, which, in turn, may lead their gap functions exhibiting a more exotic pairing symmetry, a possibility, which for \ch{NbSe_2} has garnered considerable attention in recent years, both theoretically \cite{Wickramaratne2020, Shaffer2020} and experimentally \cite{Kuzmanovi2022, Wan2023}.

A further interesting case is furnished by the results for \ch{NbSe_2} presented in Ref.\ \cite{Das2023}. These authors extract a $V_{\alpha \beta}$ using density functional theory and including electron-phonon and electron-magnon coupling $V_{\alpha \beta}^{\mathrm{ep}}$ and $V_{\alpha \beta}^{\mathrm{em}}$, respectively, such that  $V_{\alpha \beta} = V_{\alpha \beta}^{\mathrm{ep}} + \rho V_{\alpha \beta}^{\mathrm{em}}$, where $\rho$ is an adjustable parameter. Taking the numbers presented in Tables 2 and 3 of Ref. \cite{Das2023}, we have computed the eigenvalues of $V_{\alpha \beta}$ for $\rho \in [-10,10]$. In all cases, at least one, and in some cases two, eigenvalue(s) are negative. Were we to use this $V_{\alpha \beta}$ straightforwardly in a BCS approach, our constraints on the order-parameter space would have to be invoked. (The authors of Ref. \cite{Das2023} considered an Eliashberg strong-coupling approach for $N=3$. It is an interesting question under investigation how the constraints on the order-parameter space would have to be implemented in this case).

We note that the mechanism of strong interband interactions producing nonpositive eigenvalues of $V$ is quite general and holds for more general three-band interaction matrices and particularly simple systems with $N>3$, as detailed in Appendix \ref{app:interaction_matrices}. However, the limitations one must impose on the interaction matrix to ensure a given number of positive eigenvalues become increasingly more complicated as $N$ increases.
An exception to this occurs if $V$ is strictly diagonally dominant, i.e., $V_{\alpha\alpha}>\sum_{\beta \neq \alpha} |V_{\alpha\beta}| \;\forall \;\alpha$, as this prohibits any negative eigenvalues \cite{Gershgorin1931}.
Thus, if the intraband interactions are attractive and strong enough such that $V$ is strictly diagonally dominant, then nontrivial solutions to Eq.\ \eqref{eq:Gapequation} may be sought in the full order-parameter space with no further constraints. More generally, however, the interaction matrix must be checked for nonpositive eigenvalues. If there are any, the order-parameter space is constrained by Eq.\ \eqref{eq:constraints}.

\section{Conclusions}
In this paper, we have considered multiband $s$-wave BCS mean-field theories in systems with momentum-independent potentials with intra- as well as interband interactions. Such mean-field theories have been extensively used theoretically, especially following the experimental discoveries of superconducting iron pnictides, but also in recent years due to interest in superconducting monolayer TMDs with several disjoint Fermi pockets. Our main goal has been to highlight the importance of the sign of the eigenvalues of the interaction matrix and that great care must be taken to correctly identify the physically relevant solutions to the gap equation when nonpositive eigenvalues exist. By reevaluating the derivation of the multiband gap equation and treating it explicitly as a minimization of the free energy following the spirit of the original BCS paper \cite{Bardeen1957}, we have shown that in multiband systems with superconducting order parameters that do not change sign on the Fermi surface, nontrivial solutions to the gap equation which are thermodynamically unstable may exist if the interaction matrix is not positive definite. Inspired by single-band systems where the interaction between electrons must be attractive for superconducting mean-field theory to be valid, we imposed constraints on the multicomponent order parameter in such cases. 

One could consider several variations of the model presented here in future work. An obvious choice would be momentum-dependent potentials. The main challenge is then to calculate the inverse potential entering in the free energy, as it would, in general, be an inverse matrix in both momentum space and band space. Recently, determining the inverse potential in frequency space has been used to study the free energy \cite{Protter2021, Yuzbashyan2022, Zhang2022}, so it might be possible to accomplish this for momentum dependence as well. Another avenue to pursue is the renormalization of the interaction in the random phase approximation since it may affect the stability conditions presented here. Again, one then faces the challenge of calculating the inverse potential. As there are many multiband systems where the potential can be found within the random phase approximation \cite{Perali2013, Stanescu2008, Knolle2010}, it would be interesting to see if and how the constraints presented here will manifest. In the special case of separable momentum-dependent interactions, we show in Appendix \ref{app:separable-potential} that the analysis in the main text carries straightforwardly through. This extends our findings to also encompass multiband systems where the gaps may have unconventional pairing symmetries.

We have discussed in detail how strong interband interactions and repulsive intraband interactions, relevant in superconducting TMDs and iron pnictides, respectively, may necessitate introducing constraints on the $N$ superconducting order-parameter components $\{ \Delta_{\alpha} \}$. The number of constraints, which are given in Eq.\ \eqref{eq:constraints}, equals the number of nonpositive eigenvalues of the interaction matrix.
Moreover, we derived explicit expressions for the constraints in such two-band systems.
Including the constraints is imperative for the two-band theory to be physically meaningful because they ensure a lower bound on the free energy. They additionally ensure that the saddle point, which solves the gap equation in the unconstrained order-parameter space,  becomes a global energy minimum, allowing for higher-order expansions around it.
\section*{Acknowledgments}
We thank Henning Goa Hugdal, Kristian M{\ae}land, and Egor Babaev for useful discussions. This work was supported by the Research Council of Norway (RCN) through its Centres of Excellence funding scheme, Project No. 262633, ``QuSpin'', as well as RCN Project No. 323766.
\bibliography{main.bib}

\appendix
\section{Eigenvalue analysis of $N>2$ interaction matrices}\label{app:interaction_matrices}
In the main text, we argued that strong interband interactions are detrimental to a full spectrum of positive eigenvalues of the interaction matrix $V$ for $s$-wave superconductors, and that nonpositive eigenvalues necessitated invoking constraints Eq.\ \eqref{eq:constraints} on the components $\Delta_\alpha$ of the superconducting order parameter. We showed this in detail for $N=2$ bands in Sec.\ \ref{sec:2_bands} and for a specific form of $V$ for $N=3$ in Sec.\ \ref{sec:TMD}. For completeness, we write the eigenvalues of the matrix $V_3$, defined as
\begin{equation}
    V_3 = \begin{pmatrix}
        a & d & f \\
        d & a & f \\
        f & f & b
    \end{pmatrix},
    \label{symmetric_informative}
\end{equation}
such that the eigenvalues of $V_3$ are
\begin{subequations}
\label{eigenvalue_eqs}
\begin{align}
        \lambda_1 &= a - d \\
        \lambda_2 &= \frac{a + b + d}{2} + \frac{1}{2}\sqrt{(a- b + d)^2 + 8f^2} \\
        \lambda_3 &= \frac{a + b + d}{2} - \frac{1}{2}\sqrt{(a- b + d)^2 + 8f^2}.
\end{align}
\end{subequations}
Many three-band interaction matrices are in the form of $V_3$, including $V_{\mathrm{TMD}}$ in Eq.\ \eqref{V_TMD_def}. Another example is $a=b=0$, which has been studied in interband-driven superconducting iron pnictides \cite{Maiti2013, Marciani2013}.
For convenience we state explicit criteria that ensure that all eigenvalues in Eq.\ \eqref{eigenvalue_eqs} are positive
\begin{subequations}
    \begin{align}
        a &> d\\
        4(ab + bd)&>8f^2 \\
        a + b + d&>0.
    \end{align}
    \label{informative_eigvals_requirements}
\end{subequations}
If the criteria in Eq.\ \eqref{informative_eigvals_requirements} are satisfied, a system with interactions described by $V_3$ will have a well-defined superconducting mean-field theory in the full order-parameter space $\{ \Delta_{\alpha} \}, \alpha \in [1,2,3]$, with a free energy bounded from below. 

\begin{figure}
    \vspace{1mm}
    \centering
    \includegraphics[width=\linewidth,keepaspectratio]{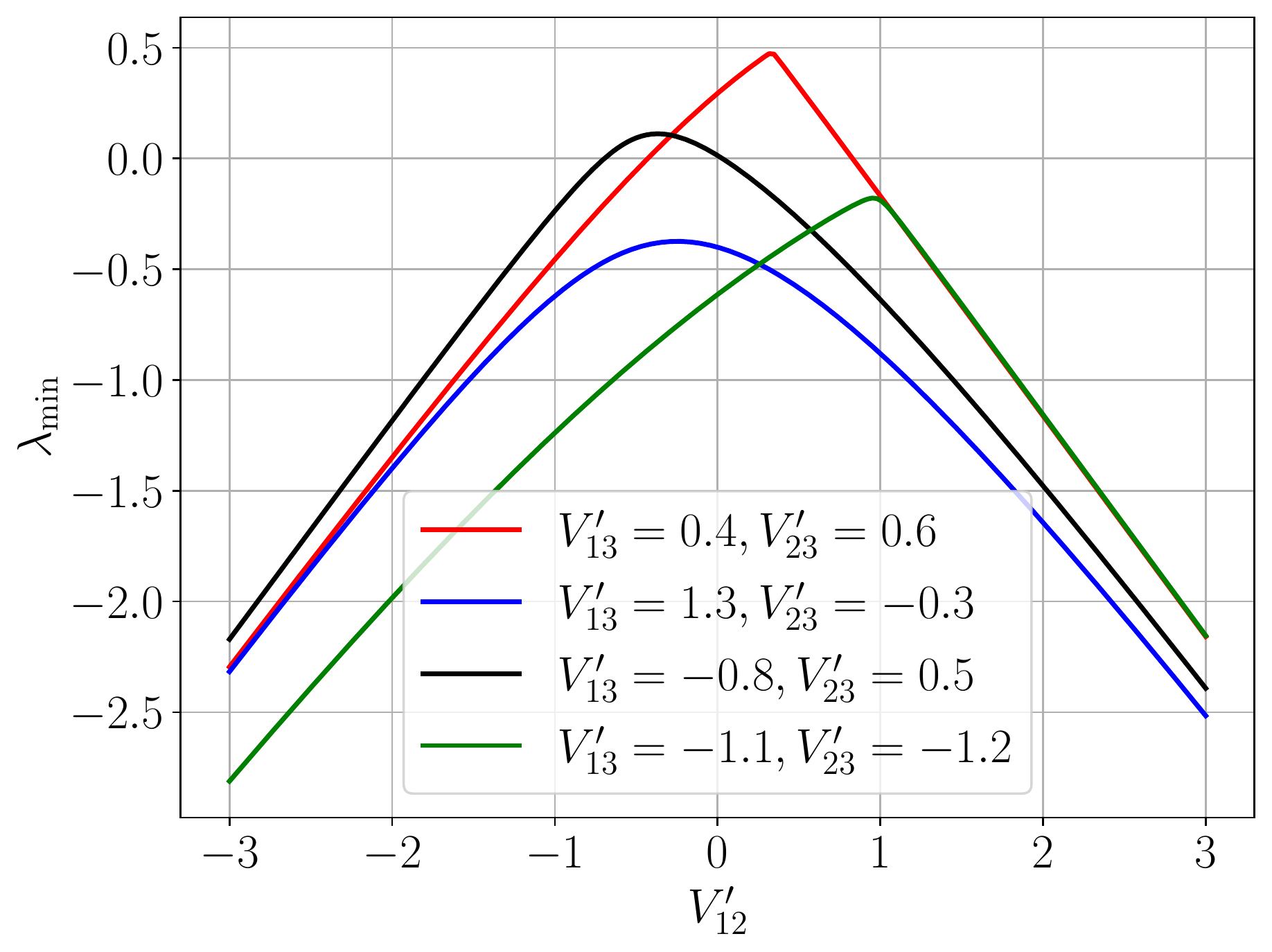}
    \caption{The smallest eigenvalue for a dimensionless three-band interaction matrix as a function of $V'_{12}$. The diagonal elements are $V'_{11} =0.7$, $V'_{22} =1$ and $V'_{33} =1.2$, and the remaining elements in the interaction matrix are different for each graph.}
    \label{fig:eigvals}
\end{figure}

To further substantiate the above claims in the case of $N=3$, we have plotted the smallest eigenvalue, $\lambda_{\mathrm{min}}$, for a more general three-band interaction matrix in Fig.\ \ref{fig:eigvals}, as a function of $V_{12}$. The matrix has attractive intraband interactions $V_{\alpha\alpha}$ of unequal strength, and each graph represents a different choice of $V_{13}$ and $V_{23}$. $\lambda_{\mathrm{min}}$ is negative for large $|V_{\alpha\beta}|$, as stated earlier. 
Unlike the two-band system, the signs of the off-diagonal matrix elements now play a more prominent role, and they can cause phase frustration as mentioned earlier. The different sign combinations of $V_{13}$ and $V_{23}$ in Fig.\ \ref{fig:eigvals} also explain why the graphs are not symmetric around $V_{12}=0$.
We further note that the complete parameter space of a three-band interaction matrix is too vast to accurately depict in one figure. However, numerical results indicate that there exists a value, denoted $V_{\alpha\beta}^c$, such that for $|V_{\alpha\beta}|> V_{\alpha\beta}^c$, at least one eigenvalue is negative, and thus neither $V$ nor $V^{-1}$ are positive definite. We find this to be the case, regardless of the other matrix elements in $V$. We also note that more general analytical expressions for the eigenvalues of three-band systems have been considered previously \cite{Yerin2013}. 

Since the complexity of the eigenvalues rapidly increases with $N$, it is in general not instructive to write the full expressions of the eigenvalues for $N=3,4$, and such expressions cannot in general be found for $N \geq 5$. To elucidate the role of strong interband interactions in such systems without turning to numerical results, we are restricted to particularly simple systems. One such system is an $N$-band system with all equal intraband and interband interactions, denoted $a$ and $b$, respectively, and one unique scattering process between two bands, denoted $c$. The interaction matrix of such a system with four bands can be written as
\begin{align}
	   V &= \begin{pmatrix}
        a & b & b & b \\
	   b & a & b & b \\
        b & b & a & c \\
        b & b & c & a
    \end{pmatrix}.
    \label{5_band_matrix}
\end{align}
Two of the eigenvalues of $V$ have simple expressions, namely $a-b$ and $a-c$, and one can immediately see that increasing either $b$ or $c$ beyond $a$ will yield negative eigenvalues. This holds also when considering a similar three or five-band system.  It therefore seems to be a quite general phenomenon that strong interband interactions necessitate that constraints be invoked on the components $\Delta_\alpha$ of the superconducting order parameter in a multiband superconductor, in order to render the theory meaningful with a free energy bounded from below.

\section{Momentum-dependent separable potential}\label{app:separable-potential}
In this appendix, we demonstrate that the impact of the eigenvalues and eigenvectors of the interaction matrix on the free energy that we have considered for momentum-independent interactions, carries straightforwardly through for the special case of a momentum-dependent separable potential, namely,
\begin{eqnarray}
    V^{\alpha \beta}_{{\bf k} {\bf k}^{\prime}} = \Lambda^{\alpha \beta}_{\eta \eta^{\prime}} ~g_{\eta {\bf k}} ~ g_{\eta^{\prime} {\bf k}^{\prime}}. 
\end{eqnarray}
Here, we have picked out one term in the more general interaction potential given in Eq. \ref{eq:interaction}, 
where $\eta$ denotes a symmetry channel that the pairing interaction has been projected onto. One example could be $g_{\eta {\bf k}}= \cos(k_x) -\cos(k_y)$, which would be relevant for describing the superconducting gap function in high-$T_c$ cuprates. (In many cases, the matrix $\Lambda^{\alpha \beta}_{\eta \eta^{\prime}}$ may be considered as dominated by its diagonal elements, $\eta = \eta^{\prime}$, but we will not assume this in the following.) 
With this in mind, we define the amplitude of the interaction as $V_{\alpha \beta} = \Lambda^{\alpha \beta}_{\eta \eta^{\prime}}$. 
We need to consider the term in the free energy given  by 
$\sum_{{\bf k} \alpha} \Delta^{\dagger}_{{\bf k} \alpha} ~b_{{\bf k} \alpha}$. We have (see main text)
\begin{eqnarray}
\Delta^{\dagger}_{{\bf k} \alpha}  = \sum_{{\bf k}^{\prime} \beta} ~ V^{\alpha \beta}_{{\bf k}{\bf k}^{\prime}}~ b^{\dagger}_{{\bf k}^{\prime} \beta}.
\end{eqnarray}
The separability of $ V^{\alpha \beta}_{{\bf k} {\bf k}^{\prime}}$ implies that we have 
\begin{eqnarray}
\Delta^{\dagger}_{{\bf k} \alpha} = \Delta_{\alpha }^\dagger ~g_{\eta {\bf k}},
\end{eqnarray}
which in turn implies that
\begin{eqnarray}
\Delta_{\alpha} = \sum_{\beta} V_{\alpha \beta} ~b_{\beta},
\label{Deltab-eq}
\end{eqnarray}
where we have defined $b_\beta \equiv \sum_{{\bf k}} g_{\eta {\bf k}} ~ b_{{\bf k} \beta}$, and $\Delta_{\alpha}$ is a gap amplitude that needs to be determined self-consistently, just as in the momentum-independent case. Thus, Eq.\ \ref{Deltab-eq} can be inverted just like in the momentum-independent case. Returning now to the pertinent term in the free energy, we have
\begin{eqnarray}
\sum_{{\bf k} \alpha} \Delta^{\dagger}_{{\bf k} \alpha} ~b_{{\bf k} \alpha} & = & \sum_{{\bf k} \alpha} \Delta^{\dagger}_{\alpha} g_{\eta {\bf k}} b_{{\bf k} \alpha} 
= \sum_{\alpha} \Delta^{\dagger}_{\alpha} ~ b_{\alpha} \nonumber \\
& = & \sum_{\alpha} \Delta^{\dagger}_{\alpha} V^{-1}_{\alpha \beta} \Delta_{\beta}.
\end{eqnarray}
Although the numerical values of the gap amplitude $\Delta_{\alpha}$, as well as the symmetry of the momentum-dependent gap $\Delta_{{\bf k} \alpha}$ will change due to the nonconstant form factors $g_{\eta \bf k}$, the boundedness of the free energy and the constraints on the space of order-parameter amplitudes, is still determined by the momentum-independent amplitudes $V_{\alpha \beta}$ of the interaction. The results of the main text would therefore apply also to multiband spin-singlet superconductors exhibiting unconventional pairing symmetry, provided that the pairing interactions are dominated by one specific symmetry channel.    

\end{document}